\documentclass[journal]{IEEEtran}
\usepackage{subfigure}
\usepackage{cite}
\usepackage{citesort}
\usepackage{amsmath,epsfig,bm}
\usepackage{pifont}
\usepackage{paralist}
\usepackage{graphicx}
\usepackage{amssymb}
\usepackage{amsmath}
\usepackage{cancel}
\usepackage[normalem]{ulem}
\usepackage[dvips]{color}
\usepackage{algorithm}
\DeclareMathOperator*{\argmax}{argmax}

\usepackage{pifont}
\usepackage{graphicx}
\usepackage{amssymb}
\usepackage{amsmath}
\usepackage{cancel}
\usepackage[normalem]{ulem}
\usepackage[dvips]{color}
\usepackage{algorithm}
\usepackage{algorithmic}

\newcommand{\Rho}{\mathrm{P}}

\newtheorem{theorem}{\noindent \textbf{Theorem}}
\newtheorem{lemma}{\noindent \textbf{Lemma}}
\newtheorem{definition}{\noindent \textbf{Definition}}

\newtheorem{remark}{\noindent \textbf{Remark}}

\begin{document}

\title{Optimization of Code Rates in SISOME \\Wiretap Channels}

\author{Shihao~Yan,
        Nan~Yang,
        Giovanni~Geraci,
        Robert~Malaney,
        and~Jinhong~Yuan
\thanks{Manuscript received Feb. 27, 2014; revised Jan. 11, 2015 and Apr. 27, 2015; accepted Jun. 22, 2015. The associate editor coordinating the review of this paper and approving it for publication was Prof. Yong Guan.}
\thanks{S. Yan, R. Malaney, and J. Yuan are with the School of Electrical Engineering and Telecommunications, The University of New South Wales, Sydney, NSW 2052, Australia (emails: shihao.yan@unsw.edu.au; r.malaney@unsw.edu.au; j.yuan@unsw.edu.au).}
\thanks{N. Yang is with the Research School of Engineering, Australia National University, Canberra, ACT 0200, Australia. He was with the School of Electrical Engineering and Telecommunications, University of New South Wales, Sydney, NSW 2052, Australia  (email: nan.yang@anu.edu.au).}
\thanks{G. Geraci is with the Wireless Networks and Decision Systems Group (WNDS), Singapore University of Technology and Design (SUTD), Singapore (email: giovanni\_geraci@sutd.edu.sg).}
\thanks{Part of this work appeared in the IEEE ICC 2014 \cite{yan2014on}.}}


\markboth{IEEE Transactions on Wireless Communications, Vol. X, No. X, Month, Year.}{Yan \MakeLowercase{\textit{et al.}}: Optimization of Code Rates in SISOME Wiretap Channels}

\vspace{-3cm}


\maketitle

\begin{abstract}
We propose a new framework for determining the wiretap code rates of single-input single-output multi-antenna eavesdropper (SISOME) wiretap channels when the capacity of the eavesdropper's channel is not available at the transmitter. In our framework we introduce the effective secrecy throughput (EST) as a new performance metric that explicitly captures the two key features of wiretap channels, namely, reliability and secrecy. Notably, the EST measures the average rate of the confidential information transmitted from the transmitter to the intended receiver without being eavesdropped on. We provide easy-to-implement methods to determine the wiretap code rates for two transmission schemes: 1) adaptive transmission scheme in which the capacity of the main channel is available at the transmitter and 2) fixed-rate transmission scheme in which the capacity of the main channel is not available at the transmitter. Such determinations are further extended into an absolute-passive eavesdropping scenario where even the average signal-to-noise ratio of the eavesdropper's channel is not available at the transmitter. Notably, our solutions for the wiretap code rates do not require us to set reliability or secrecy constraints for the transmission within wiretap channels.
\end{abstract}


\IEEEpeerreviewmaketitle

\section{Introduction}

Physical layer security for wireless communications is of
growing importance since it can guarantee the
information secrecy regardless of an eavesdropper's computational
capability and it eliminates the key distribution and
management required by traditional cryptographic techniques
\cite{Shiu,zheng2011physical,mukherjee2011robust,wang2013physical,geraci2014physical}. In the pioneering
studies \cite{shannon1949communication,wyner1975the,leung1978gaussian}, a
wiretap channel was proposed as the fundamental model to
characterize physical layer security. In the wiretap channel, an
eavesdropper (Eve) attempts to intercept the communication between a
transmitter (Alice) and an intended receiver (Bob). The prerequisite
to achieve physical layer security is that the capacity of the channel between Alice and Bob (henceforth
referred to as the main channel) is larger than the capacity of the
channel between Alice and Eve (henceforth referred to as the
eavesdropper's channel).

From the perspective of wiretap code design, the knowledge of the capacities of the main channel and the eavesdropper's channel is required at Alice in order to guarantee perfect secrecy \cite{thangaraj2007applications}. In fact, perfect
secrecy has two requirements: (i) the error probability at Bob decreases with increasing code length, and (ii) the fraction of information leakage to Eve decreases with increasing code length.
These two requirements are denoted as the reliability constraint and
the secrecy constraint, respectively \cite{thangaraj2007applications,klinc2011ldpc,harrison2013coding}. A wiretap code can be designed by choosing two code rates, namely, the codeword rate, $R_B$, and the rate of transmitted confidential information (or equivalently, the
target secrecy rate), $R_s$ \cite{thangaraj2007applications,klinc2011ldpc}. The redundancy rate, $R_E = R_B - R_s$, is used to confuse Eve. In order to
guarantee the reliability constraint of wiretap channels, the rate
of transmitted codewords has to be chosen as $R_B \leq C_B$, where $C_B$ is the capacity of the main channel. In
order to guarantee the secrecy constraint of wiretap channels, the
redundancy rate has to be chosen as $R_E > C_E$, where $C_E$ is the capacity of the eavesdropper's channel. If the values of both $C_B$ and $C_E$ are available at Alice, the
maximum target secrecy rate is achievable, which is referred to as
the secrecy capacity of a wiretap channel and is given by $C_{s} =
C_{B} - C_{E}$ \cite{gopala2008on,shafiee2009towards,khisti2010secure2,Oggier2011the}. However, the assumption that $C_E$ is available at Alice is too strong since in practice Eve may not feed back her channel state information (CSI) to Alice. In addition, in practice Alice may not know $C_B$ since Bob may not feed back $C_B$ to Alice due to the limited feedback overhead supported by the main channel\footnote{The number of bits required to feed back $C_{B}$ from Bob to Alice depends on the quantization accuracy of $C_B$.}.

We note that it is impossible for Alice to guarantee $R_E > C_E$ and fulfill the secrecy constraint in the case where only the statistical knowledge of the eavesdropper's channel is available at Alice. In this case, the
performance of wiretap channels has been characterized in terms of the ergodic secrecy capacity \cite{gopala2008on,khisti2010secure},
and in terms of the existing secrecy outage probability, i.e., $\Pr(C_s<R_s)$
\cite{parada2005secrecy,bloch2008wireless,alves2012performance,yang2012transmit,yan2013transmit}. It is important to point out that the ergodic secrecy capacity
is an average performance metric, and thus cannot be utilized to
set $R_B$ or $R_E$. The use of the existing secrecy outage probability in determining $R_B$ or $R_E$ has the drawback that it does not separate the quality of service (reliability requirements) from the secrecy requirements.

In this paper, we propose a new framework to determine the wiretap code rates when the capacity of the eavesdropper's channel is not available at Alice. Our framework is based on a new metric, referred to as the \emph{effective secrecy
throughput} (EST), which explicitly captures both the reliability constraint and the secrecy constraint of wiretap channels. The EST measures the average rate of the confidential information transmitted from Alice to Bob without being eavesdropped on. As we will discuss in more details later, the key attribute of our new metric is that it encapsulates the main features of the wiretap channel, yet can be applied to a variety of transmission schemes.

In our proposed framework we provide easy-to-implement methods to determine the wiretap code rates that achieve the locally maximum EST for two system models.
 \begin{inparaenum}[(\itshape i\upshape)]
\item The first model is a high complexity system where Bob feeds back the capacity of the main channel to Alice. We refer to the transmission scheme under this system model as the adaptive transmission scheme since Bob adaptively adjusts his wiretap code rates according to $C_B$.
\item The second model is a low complexity system where Bob does not feed back the capacity of the main channel to Alice. We refer to the transmission scheme under this system model as the fixed-rate transmission scheme since Bob has to fix his wiretap code rates when $C_B$ and $C_E$ are unavailable.
\end{inparaenum}
In the above two models, we assume that the average signal-to-noise ratio (SNR) of the eavesdropper's channel is available at Alice. In order to relax this assumption, we consider an absolute-passive eavesdropping scenario where the average SNR of the eavesdropper's channel is unavailable at Alice. For this latter scenario, we derive closed-form expressions for the average EST of the adaptive and the fixed-rate transmission schemes, based on which the wiretap code rates of these two schemes can be determined.

The rest of this paper is organized as follows. Section \ref{system_model} details the system model and the proposed new framework for determining the wiretap code rates. In Section \ref{known_CB}, the determination of the redundancy rate $R_E$ for the adaptive transmission scheme is presented. Section \ref{unknown_CB} presents the determination of $(R_B, R_E)$ for the fixed-rate transmission scheme. In Section \ref{unknown_gamma_E},  we extend the determination of the wiretap code rates into an absolute-passive eavesdropping scenario based on a proposed annulus threat model. Numerical results are provided in Section \ref{known_CB}, Section \ref{unknown_CB}, and Section \ref{unknown_gamma_E} in order to verify our analysis and provide useful insights into our solutions. Finally, Section \ref{conclusion} draws some concluding remarks.

\emph{Notation:} Scalar variables are denoted by italic symbols. Vectors and matrices are denoted by lower-case and upper-case boldface symbols, respectively. Given a complex vector $\mathbf{x}$,
$\|\mathbf{x}\|$ denotes the Euclidean norm. The $m\times{m}$ identity matrix is referred to as $\textbf{I}_{m}$ and $\mathbb{E}[\cdot]$ denotes expectation.

\section{System Model and New Framework}\label{system_model}

In this section, we detail our system model and the new framework for determining wiretap code rates ($R_B$ and $R_E$) using the proposed EST.

\subsection{System Model}

The wiretap channel of interest is where the transmitter (Alice) and the
intended receiver (Bob) are equipped with a single antenna each, and the
eavesdropper (Eve) is equipped with $N_E$ antennas. This wiretap channel is referred to as the single-input single-output multi-antenna
eavesdropper (SISOME) wiretap channel.
We note that the
deployment of multiple antennas at the eavesdropper is more conservative (from a security viewpoint) than the deployment of a single antenna at the
eavesdropper. This is due to the fact that the transmitter cannot
fully control the number of antennas at Eve.

%

We assume that
the main channel and the eavesdropper's channel are subject to
independent quasi-static Rayleigh fading with equal block length. We
also assume that Bob possesses the full knowledge of the
instantaneous CSI of the main channel, but Alice only knows the average SNR of the main channel. We further assume that Eve knows the instantaneous CSI of the
eavesdropper's channel. As such, Eve applies maximum ratio combining (MRC) \cite{shah2000performance,chen2005analysis,goldsmith2005wireless} to combine the received signals in order to exploit the
$N_{E}$-antenna diversity and maximize the probability of successful eavesdropping. This is due to the fact that when Alice transmits a single data
stream, the maximum output SNR of the eavesdropper's channel is
achieved at Eve if Eve adopts MRC. In our system
model, Alice and Bob are equipped with a single antenna each and therefore Alice transmits a single data stream to Bob.

The received signal at Bob is given by
\begin{align}\label{Bob_signal}
y_B = h x + n_B,
\end{align}
where ${h}$ is the complex coefficient of the main channel with Rayleigh fading, $x$ is the
transmit signal, and $n_B$ is the Gaussian noise of the main channel
with zero mean and variance $\sigma_{B}^2$. The transmit power
constraint is given by $\mathbb{E}[|x|^2] = P_A$, where $P_A$ is the total
transmit power. Based on \eqref{Bob_signal}, the instantaneous SNR at Bob is obtained
as
\begin{equation}\label{Bob_SNR}
\gamma_B = \frac{|h|^2 P_A}{\sigma_{B}^2},
\end{equation}
which indicates that $\gamma_B$ follows an exponential distribution with
$1/\overline{\gamma}_B$ as the rate parameter, where
$\overline{\gamma}_B=\mathbb{E}[\gamma_{B}]$.

The $N_E \times 1$ received signal vector at Eve is given
by
\begin{align}\label{Eve_signal}
\mathbf{y}_E = \mathbf{g} x + \mathbf{n}_E,
\end{align}
where $\mathbf{g}$ is the $N_E \times 1$ eavesdropper's channel
vector with independent and identically distributed (i.i.d.) Rayleigh fading entries, and $\mathbf{n}_E$ is
circularly symmetric complex Gaussian noise vector of the
eavesdropper's channel with zero mean and covariance matrix
$\mathbf{I}_{N_E}\sigma_{E}^2$. Applying MRC to exploit the
$N_{E}$-antenna diversity at Eve, the instantaneous SNR at Eve is obtained as
\begin{equation}\label{Eve_SNR}
\gamma_E=\frac{\|\mathbf{g}\|^2 P_A}{\sigma_{E}^2},
\end{equation}
which indicates that $\gamma_E$ follows a Gamma distribution with $N_E$ and
$\overline{\gamma}_E = \mathbb{E}[\gamma_{E}]/N_E$ as the shape and
scale parameters, respectively.

\subsection{New Framework for Determining Wiretap Code Rates}\label{system_model_B}

In wiretap channels, Alice intends to sent the confidential information to Bob with a high transmission rate while guaranteeing both the reliability constraint and secrecy constraint. However, if the capacity of the eavesdropper's channel is not available at Alice, the secrecy constraint cannot be guaranteed and a secrecy outage may occur. We define the secrecy outage probability as
\begin{equation}\label{secrecy_outage}
\mathcal{O}_s(R_E) = \Pr(R_E<C_E).
\end{equation}
We note that $\mathcal{O}_s(R_E)$ is different from the existing secrecy outage probability, $\Pr(C_s<R_s)$, since the latter includes not only the secrecy outage probability but also the reliability outage probability. As such, it is not exactly precise to name $\Pr(C_s < R_s)$ as the secrecy outage probability.

Likewise, if the capacity of the main channel is not available at Alice, the reliability constraint cannot be guaranteed and thus a reliability outage may occur. We define the reliability outage probability as
\begin{equation}\label{reliability_outage}
\mathcal{O}_{r}(R_B) = \Pr(R_B > C_B).
\end{equation}
Based on \eqref{reliability_outage}, we can see that the reliability constraint can be guaranteed if $C_B$ is available at Alice.

Incorporating both the secrecy outage probability and reliability outage probability, we present the EST in the following definition.
\begin{definition}
\emph{The EST (bit/channel) of a wiretap channel is defined as}
\begin{equation}\label{secrecy_throughput}
\Psi(R_B, R_E) \!=\!  \left(R_B\!-\!R_E\right)\left[1\!-\!\mathcal{O}_{r}(R_B)\right] \left[1\!-\!\mathcal{O}_{s}(R_E)\right].
\end{equation}
\end{definition}

\begin{figure}[t]
\begin{center}
{\includegraphics[width=3.0in]{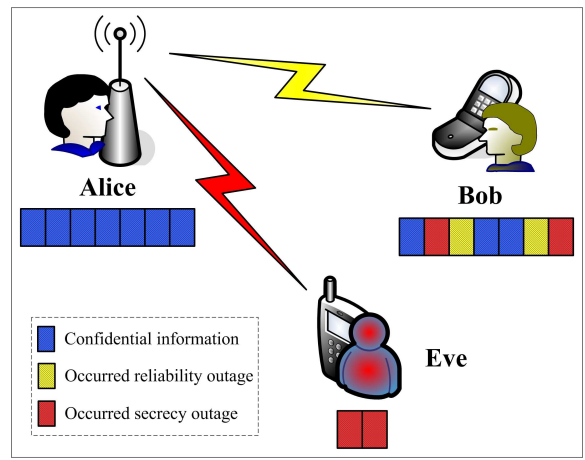}} \caption{A schematic to illustrate the operational
significance of the effective secrecy throughput (EST) over a large number of fading blocks.}
\label{fig:schematic}\end{center}
\end{figure}

In order to explicitly explain the physical meaning of the EST of a wiretap channel, we present a schematic of secure transmissions within multiple fading blocks in Fig.~\ref{fig:schematic}. In \eqref{secrecy_throughput},
$\left(R_{B}-R_{E}\right)$ quantifies the rate of transmitted
confidential information $R_{s}$ (represented by a blue block at
Alice in Fig.~\ref{fig:schematic}), while
$\left[1-\mathcal{O}_{r}\left(R_{B}\right)\right]\left[1-\mathcal{O}_{s}\left(R_{E}\right)\right]$
quantifies the probability that the information is securely
transmitted from Alice to Bob. In Fig.~\ref{fig:schematic}, the
reliability outage probability corresponds to the probability that Bob receives a yellow block, and the secrecy outage probability
corresponds to the probability that Bob receives a red block. Therefore,
the EST quantifies the average \emph{secrecy} rate at which the
messages are transmitted from Alice to Bob without being leaked to
Eve (i.e., the EST quantifies the average number of the blue blocks
achieved at Bob over a large number of fading realizations). Accordingly, the EST tells us the average rate of messages that can be securely transmitted. In the following sections, we determine values of the wiretap code rates (i.e., $R_B$, $R_E$) for the adaptive and fixed-rate transmission schemes.

Knowing the average rate of confidential messages is of
practical significance for the passive eavesdropping scenario, since
it indicates to us how much secure data can be transmitted on average
over a given period of time in a wiretap channel. We note that we cannot identify
which message is secure and which is not. We would clarify that this
is not due to the use of the EST, but is due to the adopted eavesdropping scenario. In our adopted scenario the instantaneous channel
knowledge of Eve is not available at Alice, which leads to the fact
that it is impossible to guarantee perfect secure transmission
between Alice and Bob (i.e., secrecy outage possibly occurs). Due to this, previous studies adopted the secrecy outage probability
(e.g., \cite{bloch2008wireless,yang2012transmit,he2011maximal}) and/or the throughput (e.g., \cite{zhou2011rethinking,abdallah2011keys,zhang2013on}) as the
performance metrics. We note that the average rate of
confidential messages cannot be readily evaluated by either the
secrecy outage probability or the throughput.
As such, the EST is more informative and is of more practical operational significance to Bob in some circumstances. For example, Bob may  maximize the EST without any bound  on the secrecy outage probability or the throughput.


We note here the work of \cite{zhou2011rethinking} focussed on an on-off transmission scheme for the wiretap channel and introduced a new metric that is motivated by a desire to disentangle  reliability and secrecy features. The approach used in \cite{zhou2011rethinking} is based on a new secrecy outage probability defined through a \emph{conditional} probability. This is quite different from the approach adopted here where we directly maximize a key throughput rate of the wiretap channel - the EST. Importantly, our technique does not target any transmission scheme in particular and therefore is directly applicable to a wide range of schemes and system models.

\section{Determination of Redundancy Rate for Adaptive Transmission Scheme}\label{known_CB}

In this section, we first derive a closed-form expression for the EST of the adaptive transmission scheme, based on which we provide an easy-to-implement method to determine the redundancy rate $R_E$ that achieves a locally maximum EST for this scheme. We note that applying the adaptive transmission scheme requires the capacity of the main channel to be available at Alice. As such, the complexity of the system where the adaptive transmission scheme can be applied is high since Bob has to feed back $C_B$ to Alice.

\subsection{Adaptive Transmission Scheme}

In the adaptive transmission scheme, $C_B$ is available at Alice and $R_B$ is chosen as $R_B = C_B$. As such, to design a wiretap code for the adaptive transmission scheme we only need to determine the value of $R_E$. In the adaptive transmission scheme, $R_E$ is adjusted within the
constraint $0<R_E<C_B$ according to each $\gamma_B$. Since $R_B$ is chosen as $R_B = C_B$, the reliability constraint can always be guaranteed in the adaptive transmission scheme (reliability outage probability is zero). We note that a secrecy outage may occur since $C_E$ is not available at Alice. In the following, we first present the secrecy outage probability of the adaptive transmission scheme.

As per the definition of $\mathcal{O}_{s}(R_E)$, the secrecy outage probability of the adaptive transmission scheme is
\begin{equation}\label{secrecy_outage_adaptive_rs}
\begin{split}
\mathcal{O}_{s,a}(R_E) &= \Pr(R_E < C_E) = \Pr\left(\gamma_E > 2^{R_E} - 1\right)\\
&= 1-F_{\gamma_E}\left({2^{R_E}} - 1\right),
\end{split}
\end{equation}
where $F_{\gamma_E}(\gamma_E)$ is the cumulative distribution function (cdf) of $\gamma_E$, which is
\begin{equation}\label{cdf_gamma_E_MISOME}
F_{\gamma_E}(\gamma_E) = 1 - e^{-\frac{\gamma_E}{\overline{\gamma}_E}} \sum_{j=0}^{N_E-1}\frac{1}{j!}\left(\frac{\gamma_E}{\overline{\gamma}_E}\right)^j.
\end{equation}
As such, the EST of the adaptive
transmission scheme is given in the following lemma.
\begin{lemma}\label{lemma3}
\emph{The EST of the adaptive transmission scheme is}
\begin{equation}\label{effective_secrecy_throughput_ada_MISOME}
\begin{split}
\Psi_{a}(R_E) \!=\! \left(C_B \!-\! R_E\right) \left[1 \!-\! e^{-\frac{2^{R_E}\!-\!1}{\overline{\gamma}_E}} \sum_{j=0}^{N_E-1}\frac{1}{j!}
\left(\frac{2^{R_E} \!-\! 1}{\overline{\gamma}_E}\right)^j\right].
\end{split}
\end{equation}
\end{lemma}
\begin{IEEEproof}
Since the reliability outage probability of the adaptive transmission scheme is zero and $R_B$ is set as $R_B = C_B$, based on \eqref{secrecy_throughput} the EST of the
adaptive transmission scheme is given by
\begin{equation}\label{effective_secrecy_throughput_ada}
\begin{split}
\Psi_{a}(R_E) &= \left(C_B \!-\! R_E\right) \left[1\!-\!\mathcal{O}_{s,a}(R_E)\right]\\
&= \left(C_B \!-\! R_E\right) F_{\gamma_E}\left({2^{R_E}} \!-\! 1\right).
\end{split}
\end{equation}
Substituting \eqref{cdf_gamma_E_MISOME} into
\eqref{effective_secrecy_throughput_ada}, we obtain the result in
\eqref{effective_secrecy_throughput_ada_MISOME}.
\end{IEEEproof}

\begin{figure}[!t]
    \begin{center}
    {\includegraphics[width=3.4in, height=2.8in]{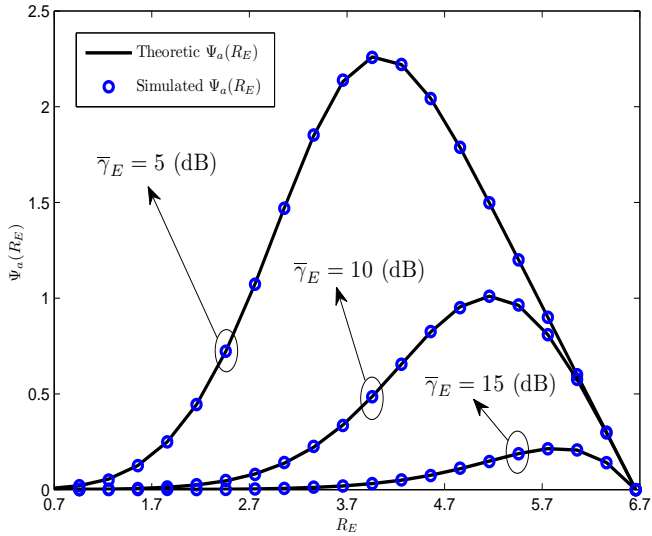}}
    \caption{EST of the adaptive transmission
    scheme, $\Psi_{a}(R_E)$, versus $R_E$ for $N_E = 3$, $\gamma_B = 20$ dB, and different values of $\overline{\gamma}_E$.}\label{fig:MISOME_secrecy_throughput_adaptive_Rs}
    \end{center}
\end{figure}

Based on the closed-form expression for $\Psi_{a}(R_E)$ given in \eqref{effective_secrecy_throughput_ada}, the value of $R_E$ that locally maximizes the EST of the adaptive transmission scheme can be numerically determined. To facilitate this numerical determination, we first derive the first-order derivative of \eqref{effective_secrecy_throughput_ada_MISOME} with respect to $R_E$ as
\begin{align}\label{effective_secrecy_rate_MISOME_d}
&\frac{\partial\Psi_{a}(R_E)}{\partial R_E} \!=\! -1 \!+\! e^{-\frac{2^{R_E} \!-\! 1}{\overline{\gamma}_E}} \sum_{j=0}^{N_E-1}\frac{1}{j!}
\left(\frac{2^{R_E} \!-\! 1}{\overline{\gamma}_E}\right)^{j}\notag\\
&~~\!+\! \left(C_B \!-\! R_E\right) \left(\frac{2^{R_E} \ln 2}{\overline{\gamma}_E}\right) \frac{ e^{-\frac{2^{R_E} \!-\! 1}{\overline{\gamma}_E}}}{(N_E\!-\!1)!}\left(\frac{2^{R_E}\!-\!1}{\overline{\gamma}_E}\right)^{N_E \!-\!1}.
\end{align}
By setting ${\partial \Psi_{a}(R_E)}/{\partial R_E} = 0$ and performing some algebraic manipulations, the value of $R_E$ that achieves a stationary point of $\Psi_{a}(R_E)$) can be obtained by solving the fixed-point equation given by
\begin{align}\label{optimal_secrecy_rate_ada_MISOME}
R_E^{\dag} &= C_B - \frac{\overline{\gamma}_E (N_E-1)!}{2^{R_E^{\dag}} \ln 2}\left(\frac{\overline{\gamma}_E}{2^{R_E^{\dag}}-1}\right)^{N_E-1}\notag\\
&~~~\times\left[e^{\frac{2^{R_E^{\dag}} - 1}{\overline{\gamma}_E}} \!-\! \sum_{j=0}^{N_E-1}\frac{1}{j!}
\left(\frac{2^{R_E^{\dag}} \!-\! 1}{\overline{\gamma}_E}\right)^j\right].
\end{align}
The second-order derivative of \eqref{effective_secrecy_throughput_ada_MISOME} with respect to $R_E$ is derived as
\begin{align}\label{adap_second_der}
\frac{\partial^2 \Psi_{a}(R_E)}{\partial R_E^2}&=\left(\frac{2^{R_E} \ln 2}{\overline{\gamma}_E}\right) \frac{ e^{-\frac{2^{R_E} \!-\! 1}{\overline{\gamma}_E}}}{(N_E\!-\!1)!}\left(\frac{2^{R_E}\!-\!1}{\overline{\gamma}_E}\right)^{N_E \!-\!1}\notag\\
&\hspace{-1cm}\times\Big[(C_B \!-\! R_E)\left(1\!-\! \frac{2^{R_E}}{\overline{\gamma}_E}\right)\ln 2\!-\!2\notag\\
&\hspace{-0.6cm}+\left(\frac{\ln 2}{2^{R_E}-1} + \ln 2 \right)(C_B \!-\! R_E)(N_E \!-\! 1)\Big].
\end{align}
An analysis of identifying stationary points via \eqref{effective_secrecy_rate_MISOME_d} and \eqref{adap_second_der} is not tractable. Instead, we investigate the nature of the stationary points via detailed simulations and numerical calculations over all the anticipated operating conditions. Although such simulations do not formally prove the globally optimal solution, we find that in the value range of $R_E$ (i.e., $0<R_E<C_B$) the stationary points obtained through \eqref{optimal_secrecy_rate_ada_MISOME}
are identified as local maxima in all simulations. An example of our simulations is shown in Fig.~\ref{fig:MISOME_secrecy_throughput_adaptive_Rs} (see numerical results section for details). As we see, for the chosen simulation parameters locally maximum operating points are easily identified.

Substituting $R_E^{\dag}$ into \eqref{effective_secrecy_throughput_ada_MISOME}, we obtain a stationary value of $\Psi_{a}(R_E)$, which is denoted by $\Psi_{a}^{\ast}$. We now conduct the asymptotic analysis of $R_E^{\dag}$ and provide some valuable insights into $R_E^{\dag}$ for $N_E = 1$ in the following remarks.

\begin{remark}\label{lemma_SNR_E_zero}
\emph{As $\overline{\gamma}_E \rightarrow 0$, we obtain $R_E^{\dag} \rightarrow 0$.}
\end{remark}
\begin{IEEEproof}
When $N_E =1$, \eqref{effective_secrecy_throughput_ada_MISOME} reduces to
\begin{equation}\label{effective_secrecy_throughput_ada_SISOSE}
\Psi_{a}(R_E) \!=\! \left( C_B\!-\!R_E \right) \left[1 \!-\! e^{-\frac{2^{R_E}\!-\!1}{\overline{\gamma}_E}}\right].
\end{equation}
It is found from (13) that $\Psi_{a}(R_E)$ converges to $(C_B\!-\!R_E)$ as $\overline{\gamma}_E
\rightarrow 0$. Therefore, due to the constraint $0<R_E<C_B$, we have $R_E^{\dag} \rightarrow 0$.
\end{IEEEproof}

It is indicated from Remark~\ref{lemma_SNR_E_zero} that Eve can be ignored if she is far from Alice. This is mainly due to the fact that the secrecy outage probability approaches zero as Eve moves further away from Alice (i.e., as the average SNR of the eavesdropper's channel
approaches zero).

\begin{remark}\label{lemma_SNR_E_infty}
\emph{As $\overline{\gamma}_E \rightarrow \infty$, $R_E^{\dag}$ can
be obtained by solving the fixed-point equation given by}
\begin{equation}\label{optimal_Rs_MISOSE_lim2}
R_E^{\dag} = C_B - \frac{1-2^{-R_E^{\dag}}}{\ln 2}.
\end{equation}
\end{remark}
\begin{IEEEproof}
Applying $\lim\limits_{x \rightarrow 0}e^{-x} \approx 1 - x$ into
\eqref{effective_secrecy_throughput_ada_SISOSE}, we obtain
\begin{equation}\label{effective_secrecy_rate_MISOSE_lim2}
\lim_{\overline{\gamma}_E \rightarrow \infty} \left( C_B\!-\!R_E \right) \left[1 \!-\! e^{-\frac{2^{R_E} \!-\! 1}{\overline{\gamma}_E}}\right] \approx \left( C_B\!-\!R_E \right) \frac{2^{R_E} \!-\! 1}{\overline{\gamma}_E}.
\end{equation}
By setting the first-order derivative of
\eqref{effective_secrecy_rate_MISOSE_lim2} with respect to $R_E$ as zero, we obtain the fixed-point equation in
\eqref{optimal_Rs_MISOSE_lim2} after some algebraic manipulations.
\end{IEEEproof}

From Remark~\ref{lemma_SNR_E_infty}, we see that $R_E^{\dag}$ does not approach $C_B$
but the effective secrecy throughput $\Psi_{a}(R_E)$ approaches zero
as $\overline{\gamma}_E \rightarrow \infty$. This is due to the fact
that the secrecy outage probability of the adaptive transmission
scheme approaches one as $\overline{\gamma}_E \rightarrow \infty$
(i.e., the secure transmission probability approaches zero as
$\overline{\gamma}_E \rightarrow \infty$). Notably, $R_E^{\dag}$
approaches a constant value that is a function of $C_B$.
This is due
to the fact that the value of $R_E$ that enables the first-order
derivative of \eqref{effective_secrecy_rate_MISOSE_lim2} to be zero does not depend on
$\overline{\gamma}_E$.

\subsection{Numerical Results}\label{numerical}

In this subsection, we present numerical results to examine the impact
of the number of antennas at Eve and the SNRs of the main channel
and eavesdropper's channel, on the redundancy rate $R_E^{\dag}$.

In Fig.~\ref{fig:MISOME_secrecy_throughput_adaptive_Rs}, we plot the
EST of the adaptive transmission scheme,
$\Psi_{a}(R_E)$, versus $R_E$ for different values of
$\overline{\gamma}_E$. The theoretic curve is obtained from
\eqref{effective_secrecy_throughput_ada_MISOME}. In this figure, we
first observe that the Monte Carlo simulations precisely match the
theoretic curves, which validates our analysis in Lemma~\ref{lemma3}. We also observe that $\Psi_{a}(R_E)$ increases as
$\overline{\gamma}_E$ decreases, which demonstrates that the worse
the eavesdropper's channel is the larger EST the adaptive transmission scheme achieves. Moreover, we
observe that a unique value of $R_E$ exists which maximizes
$\Psi_{a}(R_E)$ for a given $\gamma_B$. Focusing on the peaks of the
three curves, we also observe that $R_E^{\dag}$ decreases as
$\overline{\gamma}_E$ decreases, which demonstrates that the further
Eve is from Alice the smaller redundancy rate we set in
order to maximize the EST.

\begin{figure}[!t]
    \begin{center}
    {\includegraphics[width=3.4in, height=2.8in]{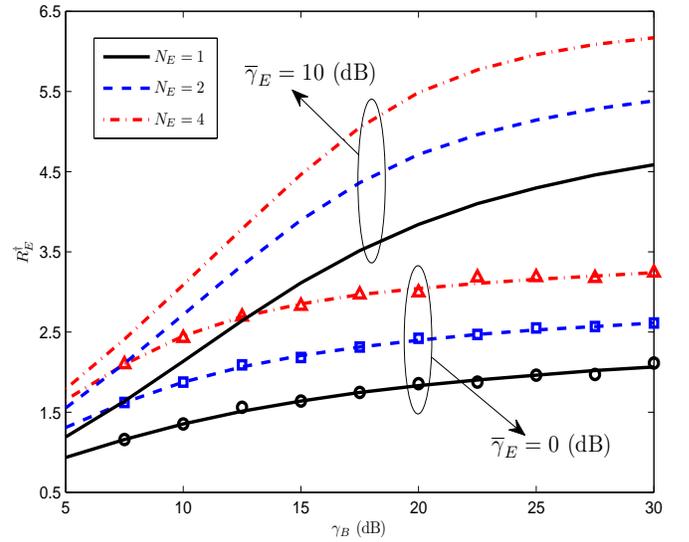}}
    \caption{Redundancy rate of the adaptive
    transmission scheme, $R_E^{\dag}$, versus ${\gamma}_B$ for different values of $N_E$ and $\overline{\gamma}_E$.}\label{fig:MISOME_always_Rs_ada}
    \end{center}
\end{figure}

In Fig.~\ref{fig:MISOME_always_Rs_ada}, we plot the redundancy rate for the adaptive transmission scheme, $R_E^{\dag}$,
versus $\gamma_B$ for different values of $N_E$ and $\overline{\gamma}_E$. The
curves represent the theoretic results for $R_E^{\dag}$ obtained
from \eqref{optimal_secrecy_rate_ada_MISOME}, and the symbols
represent the simulated results for $R_E^{\dag}$ obtained from
Monte Carlo simulations. The accuracy of \eqref{optimal_secrecy_rate_ada_MISOME} is demonstrated in this figure. As expected,
we first observe that $R_E^{\dag}$ increases as
$\overline{\gamma}_E$ increases. We also observe that $R_E^{\dag}$
first increases as $\gamma_B$ increases and then approaches a constant as $\gamma_B$ approaches large values. Furthermore, we observe that
$R_E^{\dag}$ increases as $N_E$ increases. This can be explained by the fact that a higher $N_E$ leads to a better quality of the eavesdropper's channel.

\section{Determination of Wiretap Code Rates for Fixed-Rate Transmission Scheme}\label{unknown_CB}

In this section, we first derive a closed-form expression for the EST of the fixed-rate transmission scheme, based on which we provide an easy-to-implement method to determine the codeword rate $R_B$ and redundancy rate $R_E$ that jointly achieve a locally maximum EST for this scheme. We note that the fixed-rate transmission scheme does not require the capacity of the main channel be available at Alice. As such, this scheme is a lower complexity system in which there is no feedback from Bob to Alice.

\subsection{Fixed-Rate Transmission Scheme}

In the fixed-rate transmission scheme, both $C_B$ and $C_E$ are unavailable at Alice, so we have to jointly determine $R_B$ and $R_E$ for given $\overline{\gamma}_B$ and $\overline{\gamma}_E$. Therefore, both the reliability constraint and secrecy constraint cannot be guaranteed in this scheme. In the following, we first present the reliability outage probability and secrecy outage probability.

As per the definition of $\mathcal{O}_{r}(R_B)$, the reliability outage probability of the fixed-rate transmission scheme is
\begin{equation}\label{reliability_outage_case2}
\mathcal{O}_{r,f}(R_B) = \Pr(R_B > C_B) = F_{\gamma_B}\left(2^{R_B} - 1\right),
\end{equation}
where $F_{\gamma_B}(\gamma_B)$ is the cdf of $\gamma_B$, which is
\begin{equation}\label{cdf_gamma_B_MISOME}
F_{\gamma_B}(\gamma_B)= 1 - e^{-\frac{\gamma_B}{\overline{\gamma}_B}}.
\end{equation}
According to the definition of $\mathcal{O}_s(R_E)$, the secrecy outage probability of the fixed-rate transmission scheme is
\begin{equation}\label{secrecy_outage_case2}
\mathcal{O}_{s,f}(R_E) = \Pr(R_E<C_E) = 1- F_{\gamma_E}\left(2^{R_E} - 1\right).
\end{equation}
The EST of the fixed-rate transmission scheme is presented in the following lemma.
\begin{lemma}\label{throughput_fixed_rates}
\emph{The EST of the fixed-rate transmission scheme is}
\begin{equation}\label{effective_secrecy_throughput_SIMOME}
\begin{split}
\Psi_f (R_B, R_E) &= \left(R_B-R_E\right) e^{-\frac{2^{R_B}-1}{\overline{\gamma}_B}}\\
&~~\times\left(1 \!-\! e^{-\frac{2^{R_E}\!-\!1}{\overline{\gamma}_E}} \sum_{j=0}^{N_E-1}\frac{1}{j!}\left(\frac{2^{R_E}\!-\!1}{\overline{\gamma}_E}\right)^j\right).
\end{split}
\end{equation}
\end{lemma}
\begin{IEEEproof}
As per the definition of $\Psi(R_B, R_E)$, the EST of the fixed-rate transmission scheme can be written as
\begin{equation}\label{effective_secrecy_throughput_case2}
\begin{split}
\Psi_f (R_B, R_E) &\!=\! \left(R_B\!-\!R_E\right) \left[1\!-\!\mathcal{O}_{r,f}(R_B)\right] \left[1\!-\!\mathcal{O}_{s,f}(R_E)\right]\\
&\!=\! \left(R_B\!-\!R_E\right) \left[1\!\!-\!\!F_{\gamma_B}\left(2^{R_B} \!\!-\!\! 1\right)\right] F_{\gamma_E}\left(2^{R_E} \!\!-\!\! 1\right).
\end{split}
\end{equation}
Substituting \eqref{cdf_gamma_B_MISOME} and \eqref{cdf_gamma_E_MISOME} into \eqref{effective_secrecy_throughput_case2}, we obtain the result in \eqref{effective_secrecy_throughput_SIMOME} after some algebraic manipulations.
\end{IEEEproof}

Using \eqref{effective_secrecy_throughput_case2}, the values of $(R_B, R_E)$ that achieve a locally maximum EST can be obtained through
\begin{equation}\label{numerical_fixed}
(R_B,R_E)^{\ast} = \argmax_{0 < R_B, 0 < R_E < R_B}\Psi_f (R_B, R_E).
\end{equation}
To make progress let us define two functions, $\mathcal{F}(\cdot)$ and $\mathcal{G}(\cdot)$, as
\begin{equation}
\mathcal{F}(N_E,R_E,\overline{\gamma}_E) = \sum_{j=0}^{N_E-1}\frac{1}{j!}\left(\frac{2^{R_E}-1}{\overline{\gamma}_E}\right)^j
\end{equation}
and
\begin{equation}
\mathcal{G}(N_E,R_E,\overline{\gamma}_E) = \frac{2^{R_E} \ln 2}{\overline{\gamma}_E(N_E-1)!}\left(\frac{2^{R_E} -1}{\overline{\gamma}_E}\right)^{N_E-1}.
\end{equation}
We note that $1-\mathcal{F}(N_E,R_E,\overline{\gamma}_E)e^{-\frac{2^{R_E}-1}{\overline{\gamma}_E}}$ and $e^{-\frac{2^{R_B}-1}{\overline{\gamma}_B}}$ are both positive for $0<R_E<R_B$ and finite $N_E$. Setting the first-order partial derivative of \eqref{effective_secrecy_throughput_SIMOME} with respect to $R_B$ to zero, we obtain
\begin{equation*}
1 - (R_B-R_E)\frac{2^{R_B} \ln 2}{\overline{\gamma}_B} = 0,
\end{equation*}
which results in
\begin{equation}\label{R0_SIMOME}
R_E = R_B - \frac{\overline{\gamma}_B}{2^{R_B} \ln 2}.
\end{equation}
Similarly, by setting the first-order partial derivative of \eqref{effective_secrecy_throughput_SIMOME} with respect to $R_E$ to zero, we obtain
\begin{equation*}
\mathcal{F}(N_E,R_E,\overline{\gamma}_E) - (R_B-R_E) \mathcal{G}(N_E,R_E,\overline{\gamma}_E) = e^{\frac{2^{R_E}-1}{\overline{\gamma}_E}},
\end{equation*}
which results in
\begin{equation}\label{R_SIMOME}
R_B = R_E + \frac{e^{\frac{2^{R_E}-1}{\overline{\gamma}_E}} -\mathcal{F}(N_E,R_E,\overline{\gamma}_E)}
{\mathcal{G}(N_E,R_E,\overline{\gamma}_E)}.
\end{equation}
Substituting \eqref{R0_SIMOME} into \eqref{R_SIMOME} and performing some algebraic manipulations,
the value of $R_B$ that achieves a stationary point of $\Psi_f (R_B, R_E)$
can be obtained through solving the fixed-point equation given by
\begin{equation}\label{optimal_R_SIMOME}
R_B^{\ast} = R_E^{\ast} + \frac{e^{\frac{2^{R_E^{\ast}}-1}{\overline{\gamma}_E}} -\mathcal{F}(N_E,R_E^{\ast},\overline{\gamma}_E)}
{\mathcal{G}(N_E,R_E^{\ast},\overline{\gamma}_E)},
\end{equation}
where
\begin{equation}\label{optimal_R0_SIMOME}
R_E^{\ast} = R_B^{\ast} - \frac{\overline{\gamma}_B}
{2^{R_B^{\ast}} \ln 2},
\end{equation}
and the value of $R_E$ that jointly achieves the stationary point of $\Psi_f (R_B, R_E)$ can be obtained by substituting $R_B^{\ast}$ into \eqref{optimal_R0_SIMOME}.

The Hessian matrix of \eqref{effective_secrecy_throughput_SIMOME}, which is symmetric based on Young's theorem \cite{young1984orbits}, is given by
\begin{equation}\label{hessian_matrix}
\mathbf{H} \!=\! \left[\begin{array}{cc}
\frac{\partial^2 \Psi_f (R_B, R_E)}{\partial R_B^2} & \frac{\partial^2 \Psi_f (R_B, R_E)}{\partial R_B \partial R_E} \\
\frac{\partial^2 \Psi_f (R_B, R_E)}{\partial R_E \partial R_B} & \frac{\partial^2 \Psi_f (R_B, R_E)}{\partial R_E^2} \\
\end{array}\right]
\!=\! \left[\begin{array}{cc}
\mathcal{A} & \mathcal{B} \\
\mathcal{B} & \mathcal{C} \\
\end{array}\right],
\end{equation}
where
\begin{align}
\mathcal{A} &= -\frac{2^{R_B} \ln 2}{\overline{\gamma}_B}e^{-\frac{2^{R_B}-1}{\overline{\gamma}_B}} \times \mathcal{D}\notag\\
&~~~~~~ \times \left[ 2 +  \ln 2 (R_B\!-\!R_E)\left(1-\frac{2^{R_B} }{\overline{\gamma}_B}\right)\right],\\
\mathcal{B} &= \frac{2^{R_B} \ln 2}{\overline{\gamma}_B} e^{-\frac{2^{R_B}-1}{\overline{\gamma}_B}} \times \mathcal{D}\notag \\
&~~~~~~+ e^{-\frac{2^{R_B}-1}{\overline{\gamma}_B}} \left[ 1 - (R_B-R_E)\frac{2^{R_B} \ln 2}{\overline{\gamma}_B}\right] \times \mathcal{E},\\
\mathcal{C} &= - 2 e^{-\frac{2^{R_B}-1}{\overline{\gamma}_B}} \times \mathcal{E} + (R_B\!-\!R_E)e^{-\frac{2^{R_B}-1}{\overline{\gamma}_B}}\times \mathcal{E}\notag \\
&~~~~~~ \times \left[\ln 2 + (N_E - 1)\frac{2^{R_E} \ln 2}{2^{R_E}-1 } - \frac{2^{R_E} \ln 2}{\overline{\gamma}_E}\right],
\end{align}
with
\begin{align}
\mathcal{D} &= 1 \!-\! e^{-\frac{2^{R_E}\!-\!1}{\overline{\gamma}_E}} \sum_{j=0}^{N_E-1}\frac{1}{j!}\left(\frac{2^{R_E}\!-\!1}
{\overline{\gamma}_E}\right)^j,\\
\mathcal{E} &= \frac{2^{R_E} \ln 2}{\overline{\gamma}_E (N_E-1)!}\left(\frac{2^{R_E}-1 }{\overline{\gamma}_E}\right)^{N_E-1}e^{-\frac{2^{R_E}-1}{\overline{\gamma}_E}}.
\end{align}
If $\mathcal{A} < 0$ and $\mathcal{A} \times \mathcal{C} - \mathcal{B}^2 > 0$ when $R_{B}=R_{B}^{\ast}$ and $R_{E}=R_{E}^{\ast}$, we can conclude that $R_{B}^{\ast}$ and $R_{E}^{\ast}$ lead to a locally maximum $\Psi_f (R_B, R_E)$.
However, a rigorous analysis of identifying stationary points via \eqref{optimal_R_SIMOME} and \eqref{hessian_matrix} is not tractable. Instead, we once again investigate the nature of the stationary points via detailed simulations and numerical calculations over all the anticipated operating conditions and over the rate range $0 < R_E < R_B < 100$. We find that such simulations and numerical calculations always identify the locally maximum solutions, although they do not formally identify the globally optimal solution.
One exemplary numerical result is presented in Fig.~\ref{fig:SIMOME_secrecy_throughput}, where there is a unique pair of $R_B$ and $R_E$ achieving the locally maximum $\Psi_f (R_B, R_E)$. It is highlighted that \eqref{optimal_R_SIMOME} is of great significance since it is difficult to conduct numerical search to find $R_B^{\ast}$ and $R_E^{\ast}$ under the constraint on $(R_B, R_E)$ given by $0 < R_E < R_B < + \infty$. Instead, we can solve \eqref{optimal_R_SIMOME} iteratively by setting the initial value of $R_E^{\ast}$ to zero.
Substituting $R_B^{\ast}$ and $R_E^{\ast}$ into \eqref{effective_secrecy_throughput_SIMOME}, we obtain the stationary value of $\Psi_f\left(R_B, R_E\right)$, which is denoted by $\Psi^{\ast}_f$.

We next conduct the asymptotic analysis for $N_E =1$ and offer some valuable insights into $R_B^{\ast}$ and $R_E^{\ast}$ in the following remarks.
\begin{remark}\label{corollary_case2_E}
\emph{As $\overline{\gamma}_E \rightarrow 0$, $R_B^{\ast}$ for a given $\overline{\gamma}_B$ can be obtained through solving the fixed-point equation given by}
\begin{equation}\label{optimal_R_SISOSE_limE_0}
R_B^{\ast} = \frac{\overline{\gamma}_B}{2^{R_B^{\ast}} \ln 2},
\end{equation}
\emph{and the corresponding $R_E^{\ast}$ approaches zero.}
\end{remark}
\begin{IEEEproof}
As $\overline{\gamma}_E \rightarrow 0$, we obtain $\mathcal{O}_{s,f}(R_E) \rightarrow 0$. Accordingly, \eqref{effective_secrecy_throughput_SIMOME} with $N_E =1$ reduces to
\begin{equation}\label{effective_secrecy_throughput_SISOSE_lim0}
\Psi_f (R_B, R_E) = \left(R_B-R_E\right) e^{-\frac{2^{R_B}-1}{\overline{\gamma}_B}}.
\end{equation}
Setting the first-order derivative of \eqref{effective_secrecy_throughput_SISOSE_lim0} with respect to $R_B$ as zero, we obtain the result in \eqref{optimal_R_SISOSE_limE_0} after some algebraic manipulations.
\end{IEEEproof}

From Remark~\ref{corollary_case2_E}, we see that Eve can be ignored if she is far from
Alice since $R_E^{\ast} \rightarrow 0$ as $\overline{\gamma}_E
\rightarrow 0$. This is due to the fact that the secrecy outage
probability of the fixed-rate transmission scheme,
$\mathcal{O}_{s,f}(R_E)$, approaches zero as $\overline{\gamma}_E
\rightarrow 0$. We also note that $R_B^{\ast}$ is a function of
$\overline{\gamma}_B$ only, which can be explained by the fact that
$R_B^{\ast}$ is determined through maximizing $R_B \left[1 -
\mathcal{O}_{r,f}(R_B)\right]$. We further note that $\Psi_f^{\ast}$
approaches a constant value which is determined by
$\overline{\gamma}_B$ as $\overline{\gamma}_E \rightarrow 0$, due to
that $\mathcal{O}_{s,f}(R_E)\rightarrow 0$ as $\overline{\gamma}_E
\rightarrow 0$.

\begin{remark}\label{corollary_case2_E_infty}
\emph{As $\overline{\gamma}_E \rightarrow \infty$, $R_B^{\ast}$ for a given $\overline{\gamma}_B$ can be obtained through solving the fixed-point equation given by}
\begin{equation}\label{optimal_R_SISOSE_limE_infty}
R_B^{\ast} = R_E^{\ast} + \frac{2^{R_E^{\ast}}-1}
{2^{R_E^{\ast}} \ln 2},
\end{equation}
\emph{where}
\begin{equation}\label{optimal_R0_SISOSE_limE_infty}
R_E^{\ast} = R_B^{\ast} - \frac{\overline{\gamma}_B}
{2^{R_B^{\ast}} \ln 2},
\end{equation}
\emph{and $R_E^{\ast}$ can be obtained by substituting $R_B^{\ast}$ into \eqref{optimal_R0_SISOSE_limE_infty}.}
\end{remark}
\begin{IEEEproof}
When $N_E = 1$, \eqref{effective_secrecy_throughput_SIMOME} reduces to
\begin{equation}\label{effective_secrecy_throughput_SISOSE}
\Psi_f (R_B, R_E) = \left(R_B\!-\!R_E\right) e^{-\frac{2^{R_B}-1}{\overline{\gamma}_B}} \left(1 \!-\! e^{-\frac{2^{R_E}\!-\!1}{\overline{\gamma}_E}}\right).
\end{equation}
Since $R_E$ is still finite as $\overline{\gamma}_E \rightarrow \infty$, we apply $\lim\limits_{x \rightarrow 0}e^{-x} \approx 1 - x$ into \eqref{effective_secrecy_throughput_SISOSE} and obtain
\begin{align}\label{reliable_secrecy_outage_SISOSE_limE}
\lim_{\overline{\gamma}_E \!\rightarrow\! \infty}\Psi_f\left(R_B, R_E\right) \!\approx\! \left(R_B-R_E\right) e^{-\frac{2^{R_B} - 1}{\overline{\gamma}_B}}\frac{2^{R_E}-1}{\overline{\gamma}_E}.
\end{align}
We note $\frac{2^{R_E}-1}{\overline{\gamma}_E} > 0$ due to $R_E > 0$ as $\overline{\gamma}_E \rightarrow \infty$. Setting the first-order derivative of \eqref{reliable_secrecy_outage_SISOSE_limE} with respect to $R_B$ as zero, we obtain
\begin{equation}\label{reliable_secrecy_outage_SISOSE_limE_0_dR}
\left(R_B-R_E\right)\frac{2^{R_B} \ln 2}{\overline{\gamma}_B} = 1.
\end{equation}
Likewise, setting the first-order derivative of \eqref{reliable_secrecy_outage_SISOSE_limE} with respect to $R_E$ as zero, we obtain
\begin{equation}\label{reliable_secrecy_outage_SISOSE_limE_dR0}
-\frac{2^{R_E}-1}{\overline{\gamma}_E} + \left(R_B-R_E\right) \frac{2^{R_E} \ln 2}{\overline{\gamma}_E} = 0,
\end{equation}
which results in (due to $\overline{\gamma}_E \neq 0$)
\begin{equation}\label{reliable_secrecy_outage_SISOSE_limE_dR0_f}
1- 2^{R_E} + \left(R_B-R_E\right){2^{R_E} \ln 2} = 0.
\end{equation}
Substituting \eqref{reliable_secrecy_outage_SISOSE_limE_0_dR} into \eqref{reliable_secrecy_outage_SISOSE_limE_dR0_f}, we obtain the fixed-point equation in \eqref{optimal_R_SISOSE_limE_infty} after some algebraic manipulations.
\end{IEEEproof}

It is highlighted from Remark~\ref{corollary_case2_E_infty} that \eqref{optimal_R_SISOSE_limE_infty} and  \eqref{optimal_R0_SISOSE_limE_infty} are
independent of $\overline{\gamma}_E$, which indicates that
$R_B^{\ast}$ and $R_E^{\ast}$ are not functions of
$\overline{\gamma}_E$ as $\overline{\gamma}_E \rightarrow \infty$.
This is due to the fact that the secrecy outage probability
$\mathcal{O}_{s,f}(R_E)$ approaches 1 as $\overline{\gamma}_E
\rightarrow \infty$.

\subsection{Numerical Results}

In this subsection, we present numerical results to examine the impact
of the number of antennas at Eve and the SNRs of the main channel
and eavesdropper's channel, on $R_B^{\ast}$ and $R_E^{\ast}$. We also conduct a thorough performance comparison
between the adaptive transmission
scheme and the fixed-rate transmission scheme in terms of the EST.

\begin{figure}[!t]
    \begin{center}
    {\includegraphics[width=3.3in,height=2.8in]{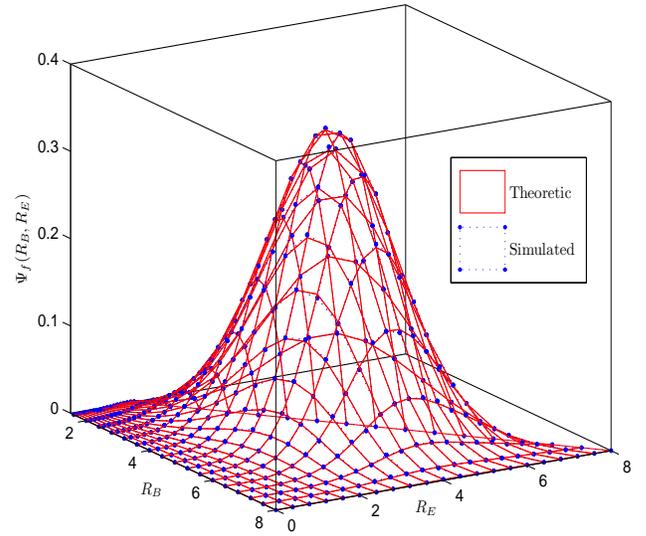}}
    \end{center}
    \caption{EST of the fixed-rate transmission scheme, $\Psi_f(R_B, R_E)$, versus $R_B$ and $R_E$ for $N_E = 3$, $\overline{\gamma}_B = 15$ dB, and $\overline{\gamma}_E = 5$ dB.}\label{fig:SIMOME_secrecy_throughput}
\end{figure}

In Fig.~\ref{fig:SIMOME_secrecy_throughput}, we plot the EST of the fixed-rate transmission scheme, $\Psi_f(R_B, R_E)$, versus $R_B$ and $R_E$. The theoretic $\Psi_f(R_B, R_E)$ curve is generated via \eqref{effective_secrecy_throughput_SIMOME}. In this figure, we first observe that the Monte Carlo simulation result precisely matches the theoretic curve. Moreover, we observe that there is indeed a unique pair of $R_B$ and $R_E$ that achieves a locally maximum value of $\Psi_f(R_B, R_E)$. This demonstrates that we can determine the values of $(R_B, R_E)$ that achieve a locally maximum $\Psi_f(R_B, R_E)$ based on our proposed framework.

\begin{figure}[!t]
    \begin{center}
    {\includegraphics[width=3.3in,height=2.8in]{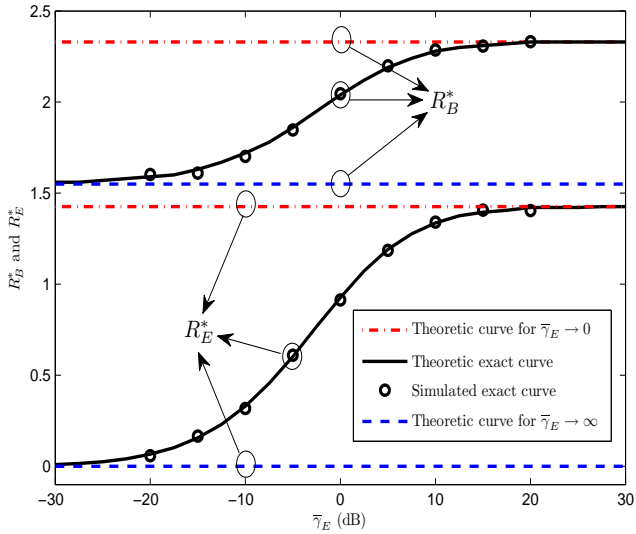}}
    \end{center}
    \caption{Wiretap code rates of the fixed-rate transmission scheme, $R_B^{\ast}$ and $R_E^{\ast}$, versus $\overline{\gamma}_E$ for $N_E = 1$ and $\overline{\gamma}_B = 5$ dB.} \label{fig:optimal_rates_SISOSE_limE}
\end{figure}

In Fig.~\ref{fig:optimal_rates_SISOSE_limE}, we plot the wiretap code rates of the fixed-rate transmission scheme, $R_B^{\ast}$ and $R_E^{\ast}$, versus $\overline{\gamma}_E$. The exact curves of $R_B^{\ast}$ and $R_E^{\ast}$ are obtained by
solving \eqref{optimal_R_SIMOME} and \eqref{optimal_R0_SIMOME}, respectively. The curve of $R_B^{\ast}$ for $\overline{\gamma}_E \rightarrow 0$ is generated by solving \eqref{optimal_R_SISOSE_limE_0}, and $R_E^{\ast}$ for $\overline{\gamma}_E \rightarrow 0$ is approximated as zero. The curves of $R_B^{\ast}$ and $R_E^{\ast}$ for $\overline{\gamma}_E \rightarrow \infty$ are achieved by solving \eqref{optimal_R_SISOSE_limE_infty} and \eqref{optimal_R0_SISOSE_limE_infty}, respectively. As we have checked, in Fig.~\ref{fig:optimal_rates_SISOSE_limE} $R_B^{\ast}$ and $R_E^{\ast}$ jointly achieve a locally maximum $\Psi_f(R_B, R_E)$.
In this figure, we first observe that the Monte Carlo simulated $R_B^{\ast}$ and $R_E^{\ast}$ precisely match the theoretic $R_B^{\ast}$ and $R_E^{\ast}$, respectively.
We also observe that the exact curves of $R_B^{\ast}$ and $R_E^{\ast}$ approach the asymptotic curves of $R_B^{\ast}$ and $R_E^{\ast}$, respectively, as $\overline{\gamma}_E \rightarrow 0$ and $\overline{\gamma}_E \rightarrow \infty$. This observation confirms the accuracy of our asymptotic analysis given in Remark~\ref{corollary_case2_E} and Remark~\ref{corollary_case2_E_infty}. Finally, we observe that both $R_B^{\ast}$ and $R_E^{\ast}$ increase as $\overline{\gamma}_E$ increases, but ($R_B^{\ast}-R_E^{\ast}$) decreases as $\overline{\gamma}_E$ increases.


\begin{figure}[!t]
    \begin{center}
    {\includegraphics[width=3.3in,height=2.8in]{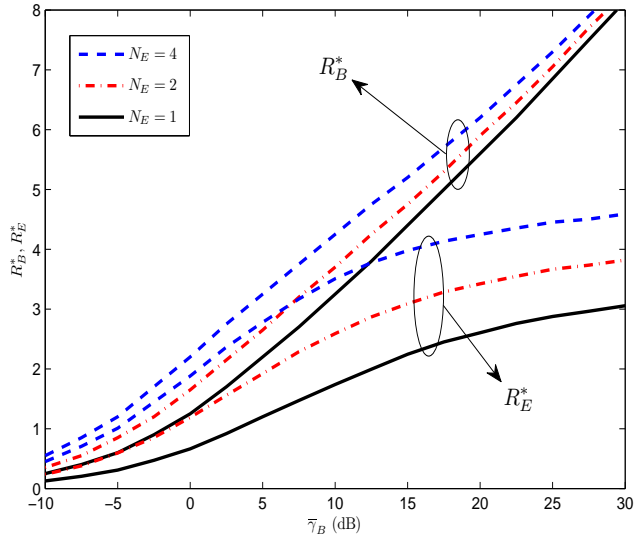}}
    \end{center}
    \caption{Wiretap code rates of the fixed-rate transmission scheme, $R_B^{\ast}$ and $R_E^{\ast}$, versus $\overline{\gamma}_B$ for $\overline{\gamma}_E = 5$ dB.} \label{fig:optimal_rates_SISOSE_limE_B}
\end{figure}

In Fig.~\ref{fig:optimal_rates_SISOSE_limE_B}, we plot $R_B^{\ast}$ and $R_E^{\ast}$ versus $\overline{\gamma}_B$ for different values of $N_E$, which have been confirmed as the values of $R_B$ and $R_E$ that jointly achieve a locally maximum $\Psi_f(R_B, R_E)$.
In this figure, we first observe that both $R_B^{\ast}$ and $R_E^{\ast}$ increase as $\overline{\gamma}_B$ increases, and ($R_B^{\ast} - R_E^{\ast}$) increases as $\overline{\gamma}_B$ increases. We also observe that as $\overline{\gamma}_B \rightarrow 0$ both $R_B^{\ast}$ and $R_E^{\ast}$ approach zero, which indicates that a positive EST cannot be achieved when Bob is very far from Alice. Furthermore, we observe that $R_B^{\ast}$ is still a function of $\overline{\gamma}_B$ as $\overline{\gamma}_B \rightarrow \infty$, but $R_E^{\ast}$ approaches a specific constant value. Finally, we observe that both $R_B^{\ast}$ and $R_E^{\ast}$ increase as $N_E$ increases, but ($R_B^{\ast} - R_E^{\ast}$) decreases as $N_E$ increases.

\begin{figure}[!t]
    \begin{center}
    {\includegraphics[width=3.3in, height=2.8in]{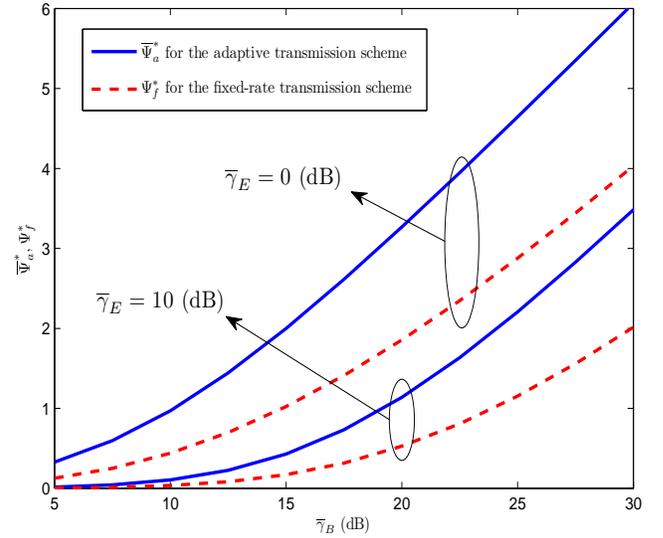}}
    \end{center}
    \caption{$\overline{\Psi}_{a}^{\ast}$ of the adaptive transmission scheme and $\Psi_f^{\ast}$ of fixed-rate transmission scheme versus $\overline{\gamma}_B$ for $N_E = 2$.}\label{fig:SISOME_fix_ada_unknown_com}
\end{figure}

Now, we conduct a thorough comparison between the adaptive transmission scheme and the fixed-rate transmission scheme.
The results are presented in Fig.~\ref{fig:SISOME_fix_ada_unknown_com}, where $\overline{\Psi}_{a}^{\ast}$ is the average locally maximum
EST of the adaptive transmission scheme,
obtained by
$\overline{\Psi}_{a}^{\ast}=\mathbb{E}_{\gamma_B}\left[\Psi_{a}
^{\ast}\right]$. We first observe that both $\overline{\Psi}_{a}^{\ast}$ and $\Psi_f^{\ast}$ increase as $\overline{\gamma}_B$ increases, but decrease as $\overline{\gamma}_E$ increases. This indicates that the locations of Alice, Bob, and Eve are of importance in wiretap channels. As expected, we observe that the adaptive transmission scheme achieves higher EST than the fixed-rate transmission scheme. In addition, we observe that the EST gain of the adaptive transmission scheme over the fixed-rate transmission scheme is negligible in the regime of low $\overline{\gamma}_B$ (relative to $\overline{\gamma}_E$), but significant in the regime of high $\overline{\gamma}_B$.
Of course, the EST is enhanced at the cost of feeding back $C_B$ to Alice and adjusting $R_B$ and $R_E$ for each realization of the main channel.

\section{Determination of Wiretap Code Rates Within An Absolute-Passive Eavesdropping Scenario}\label{unknown_gamma_E}

In this section, we extend the determination of the wiretap code rates for the adaptive and fixed-rate transmission schemes into an absolute-passive eavesdropping scenario, where no SNR information of the eavesdropper's channel (not even the average value) is made available to Alice. To relax the assumption in the previous two sections that $\overline{\gamma}_E$ is known at Alice, we propose a new threat model, referred to as the annulus threat model. For this threat model, we derive closed-form expressions for the average EST of the adaptive and fixed-rate transmission schemes, based on which the wiretap code rates can be determined.

\subsection{Annulus Threat Model}

\begin{figure}[!t]
    \begin{center}
    {\includegraphics[width=3.1in,height=2.7in]{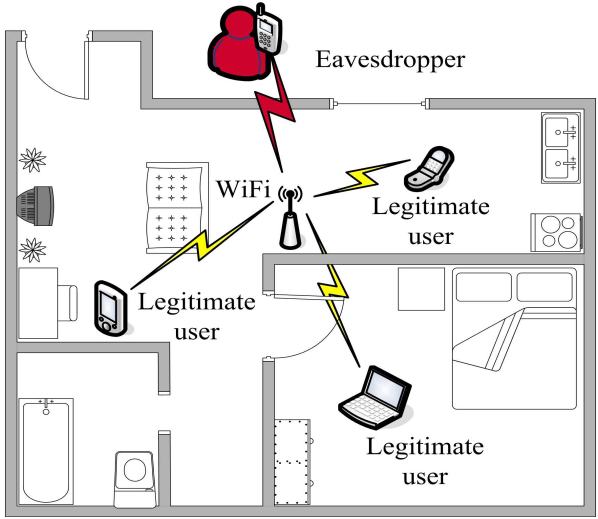}}
    \end{center}
    \caption{Illustration of a practical scenario based on which the annulus threat model is proposed.}\label{fig:threat}
\end{figure}

Fig.~\ref{fig:threat} depicts a practical scenario where physical layer security may apply. In this scenario, the Wi-Fi point (Alice) and the legitimate user (Bob) are located inside a property (e.g., a house), but the eavesdropper (Eve) is bounded outside the property. In practice, Eve cannot be infinitely far from Alice. As such, in practical wiretap channels the distance between Alice and Eve should be larger than a specific value and less than another specific value. This motivates us to propose an annulus threat model. In the annulus threat model, we assume that Eve's location is uniformly distributed inside an annulus bounded by two concentric circles, where $\rho_i$ and $\rho_o$ are the radii of the inner circle and the outer circle, respectively, and Alice is at the center of the two concentric circles.

In our annulus threat model, we denote the distance between Alice and Eve as $\rho$. Based on the path loss model, the average SNR of the eavesdropper's channel is a function of $\rho$, which can be expressed as \cite{goldsmith2005wireless}
\begin{equation}\label{path loss}
\overline{\gamma}_E = c_0 \rho^{-\eta},
\end{equation}
where $c_0 = {\overline{\gamma}_0}/{\rho_r^{-\eta}}$, $\overline{\gamma}_0$ is the reference average SNR of the eavesdropper's channel at the reference distance $\rho_r$, and $\eta$ is the path loss exponent. In the previous sections, the EST is derived as a function of $\overline{\gamma}_E$. In this section, we derive the average secrecy throughput over $\overline{\gamma}_E$ under the annulus threat model. To this end, we first present the probability density function (pdf) of $\rho^2$ in the following lemma.
\begin{lemma}
\emph{The square of the distance between Alice and Eve, $\rho^2$, follows a uniform distribution with $\rho_i^2$ and $\rho_o^2$ as the lower bound and upper bound, respectively, i.e., $\rho^2 \sim \mathbf{U}(\rho_i^2, \rho_o^2)$.}
\end{lemma}
\begin{IEEEproof}
See Appendix.
\end{IEEEproof}

\subsection{Adaptive Transmission Scheme}

We note that the adaptive transmission scheme represents a high complexity system where Bob feeds back $C_B$ to Alice, and we only need to determine the value of $R_E$ for this scheme since $R_B$ is set as $R_B = C_B$. We derive the average secrecy throughput of the adaptive transmission scheme for the annulus threat model in the following theorem.
\begin{theorem}
\emph{The average EST of the adaptive transmission scheme for the annulus threat model is}
\begin{equation}\label{throughput_ada_threat}
\begin{split}
\widetilde{\Psi}_{a}(R_{E},&\rho_{i},\rho_{o}) = \left(C_B \!-\! R_E\right)\\
&\times \left(1 \!\!-\!\! \sum_{j=0}^{N_E-1}\frac{2u^j}{j!}\frac{\gamma(v, u \rho_o^{\eta}) \!-\! \gamma(v, u \rho_i^{\eta})}{(\rho_o^2 \!-\! \rho_i^2)\eta u^v}\right),
\end{split}
\end{equation}
\emph{where $u = \left({2^{R_E}-1}\right)/{c_0}$, $v = j + 2/\eta$, and $\gamma(\cdot, \cdot)$ is the incomplete gamma function \cite[Eq. (8.350.1)]{gradshteuin2007table}.}
\begin{IEEEproof}
Under the annulus threat model, the average EST of the adaptive transmission scheme is
\begin{equation}\label{proof4_1}
\widetilde{\Psi}_{a}(R_{E},\rho_{i},\rho_{o}) = \mathbb{E}_{\rho}[\Psi_{a}(R_E)] = \int_{\rho_i^2}^{\rho_o^2} \frac{\Psi_{a}(R_E)}{\rho_o^2-\rho_i^2} d \rho^2.
\end{equation}
Substituting \eqref{effective_secrecy_throughput_ada_MISOME} and \eqref{path loss} into \eqref{proof4_1}, we obtain
\begin{equation}\label{average_ada}
\begin{split}
&\widetilde{\Psi}_{a}(R_{E},\rho_{i},\rho_{o}) = \int_{\rho_i^2}^{\rho_o^2} \frac{\Psi_{a}(R_E)}{\rho_o^2-\rho_i^2} d \rho^2\\
&= \frac{\left(C_B \!-\! R_E\right)}{\rho_o^2 \!-\! \rho_i^2}\int_{\rho_i^2}^{\rho_o^2}\left[1 \!-\! e^{-\frac{2^{R_E}\!-\!1}{c_0 \rho^{-\eta}}} \sum_{j=0}^{N_E-1}\frac{1}{j!}
\left(\frac{2^{R_E} \!-\! 1}{c_0 \rho^{-\eta}}\right)^j\right] d \rho^2\\
&= \left(C_B \!-\! R_E\right) - \frac{\left(C_B \!-\! R_E\right)}{\rho_o^2 - \rho_i^2}\sum_{j=0}^{N_E-1}\frac{1}{j!}
\left(\frac{2^{R_E} \!-\! 1}{c_0}\right)^j\\
&~~~~\times \left[\int_{0}^{\rho_o^2} \rho^{j \eta} e^{-\frac{2^{R_E}\!-\!1}{c_0}\rho^{\eta}}  d \rho^2 \!-\! \int_{0}^{\rho_i^2} \rho^{j \eta} e^{-\frac{2^{R_E}\!-\!1}{c_0}\rho^{\eta}}  d \rho^2\right].
\end{split}
\end{equation}
We solve the integrals in \eqref{average_ada} with the aid of \cite[Eq. (3.381.8)]{gradshteuin2007table}, and obtain the desired result in \eqref{throughput_ada_threat} after some algebraic manipulations.
\end{IEEEproof}
\end{theorem}

\begin{figure}[!t]
\begin{center}
{\includegraphics[width=3.4in,height=2.8in]{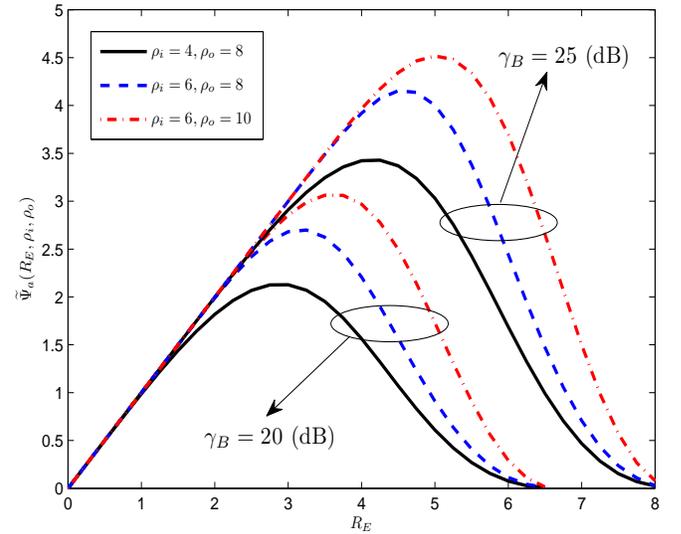}}
\caption{Average EST of the adaptive transmission scheme for the annulus threat model versus $R_E$ for $N_E = 2$, $\overline{\gamma}_0 = 30$ dB, $\rho_r = 1$, and $\eta = 3$.}
\label{fig:ada_concave}\end{center}
\end{figure}

The value of $R_E$ that achieves a locally maximum $\widetilde{\Psi}_{a}(R_{E},\rho_{i},\rho_{o})$ can be determined through
\begin{equation}
\widetilde{R}_E^{\dag} = \argmax_{0 < R_E < C_B}\widetilde{\Psi}_{a}(R_{E},\rho_{i},\rho_{o}).
\end{equation}
One exemplary numerical result is presented in Fig.~\ref{fig:ada_concave}, which shows that there is a unique value of $R_E$ within $0<R_E<C_B$ achieving the locally maximum $\widetilde{\Psi}_{a}(R_{E},\rho_{i},\rho_{o})$. Substituting $\widetilde{R}_E^{\dag}$ into \eqref{throughput_ada_threat}, we obtain the locally maximum value of $\widetilde{\Psi}_{a}(R_{E},\rho_{i},\rho_{o})$, denoted by $\widetilde{\Psi}^{\ast}_a(\rho_i, \rho_o)$.

\subsection{Fixed-Rate Transmission Scheme}

We reminder the reader that the fixed-rate transmission scheme represents a lower complexity system where Bob does not feed back $C_B$ to Alice, and we have to jointly determine the values of $R_B$ and $R_E$. We derive the average EST of the fixed-rate transmission scheme for the annulus threat model in the following theorem.

\begin{figure}[!t]
\begin{center}
{\includegraphics[width=3.3in, height=2.8in]{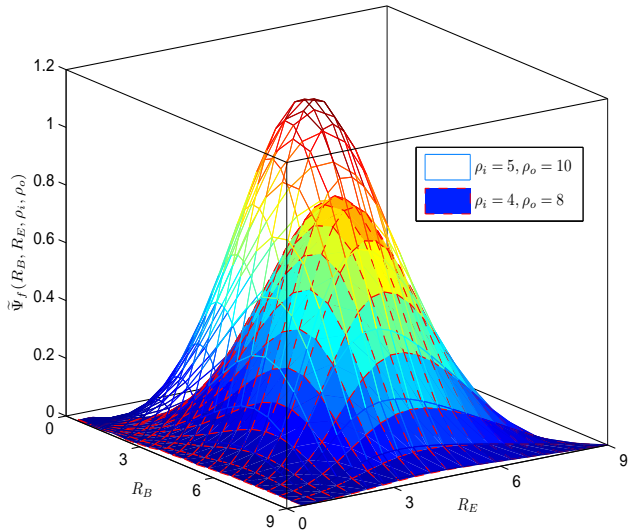}}
\caption{Average EST of the fixed-rate transmission scheme for the annulus threat model versus $R_B$ and $R_E$ for $N_E = 2$, $\overline{\gamma}_B = 20$dB, $\overline{\gamma}_0 = 30$ dB, $\rho_r = 1$, and $\eta = 3$.}
\label{fig:fixed_concave}\end{center}
\end{figure}

\begin{theorem}
\emph{The average EST of the fixed-rate transmission scheme for the annulus threat model is}
\begin{equation}\label{throughput_fixed_threat}
\begin{split}
\widetilde{\Psi}_f(&R_B, R_E, \rho_i, \rho_o) ={\left(R_B-R_E\right)} e^{-\frac{2^{R_B}-1}{\overline{\gamma}_B}}\\
&~~~~\times \left(1 - \sum_{j=0}^{N_E-1}\frac{2w^j}{j!}\frac{\gamma(v, w \rho_o^{\eta}) \!-\! \gamma(v, w \rho_i^{\eta})}{({\rho_o^2 \!-\! \rho_i^2})\eta w^v}\right),
\end{split}
\end{equation}
\emph{where $w = \left({2^{R_E}-1}\right)/{c_0}$}.
\begin{IEEEproof}
Under the annulus threat model, the average EST of the fixed-rate transmission scheme is
\begin{equation}\label{proof5_1}
\begin{split}
\widetilde{\Psi}_f(R_B, R_E, \rho_i, \rho_o) &= \mathbb{E}_{\rho}[\Psi_{f}(R_B, R_E)]\\
&= \int_{\rho_i^2}^{\rho_o^2} \frac{\Psi_{f}(R_B, R_E)}{\rho_o^2-\rho_i^2} d \rho^2.
\end{split}
\end{equation}
Substituting \eqref{effective_secrecy_throughput_SIMOME} and \eqref{path loss} into \eqref{proof5_1}, we obtain
\begin{align}
&\widetilde{\Psi}_f(R_B, R_E, \rho_i, \rho_o) = \int_{\rho_i^2}^{\rho_o^2} \frac{\Psi_{f}(R_B, R_E)}{\rho_o^2-\rho_i^2} d \rho^2\notag
\end{align}
\begin{align}\label{average_fix}
&=\frac{\left(R_B-R_E\right)}{\rho_o^2 \!-\! \rho_i^2} e^{-\frac{2^{R_B}-1}{\overline{\gamma}_B}}\notag \\
&~~~~\times \int_{\rho_i^2}^{\rho_o^2} \left[1 \!-\! e^{-\frac{2^{R_E}\!-\!1}{c_0 \rho^{-\eta}}} \sum_{j=0}^{N_E-1}\frac{1}{j!}\left(\frac{2^{R_E}\!-\!1}
{c_0 \rho^{-\eta}}\right)^j\right]d \rho^2 \notag\\
&={\left(R_B-R_E\right)} e^{-\frac{2^{R_B}-1}{\overline{\gamma}_B}}\notag\\
&~~~~\times \bigg[1 - \frac{1}{\rho_o^2 \!-\! \rho_i^2}\sum_{j=0}^{N_E-1}\frac{1}{j!}\left(\frac{2^{R_E}\!-\!1}
{c_0}\right)^j\bigg.\notag\\
&~~~~~~\times\left.\left(
\int_{0}^{\rho_o^2} \rho^{j \eta} e^{-\frac{2^{R_E}\!-\!1}{c_0}\rho^{\eta}}  d \rho^2 \!-\! \int_{0}^{\rho_i^2} \rho^{j \eta} e^{-\frac{2^{R_E}\!-\!1}{c_0}\rho^{\eta}}  d \rho^2\right)\right].
\end{align}
We solve the integrals in \eqref{average_fix} with the aid of \cite[Eq. (3.381.8)]{gradshteuin2007table}, and obtain the result in \eqref{throughput_fixed_threat} after some algebraic manipulations.
\end{IEEEproof}
\end{theorem}

The values of $(R_B, R_E)$ that achieve a locally maximum $\widetilde{\Psi}_f(R_B, R_E, \rho_i, \rho_o)$ can be determined through
\begin{equation}
(R_B, R_E)^{\ast}_{\rho} = \argmax_{0 < R_B, 0<R_E<R_B}\widetilde{\Psi}_f(R_B, R_E, \rho_i, \rho_o).
\end{equation}
The values of $R_B$ and $R_E$ in $(R_B, R_E)^{\ast}_{\rho}$ are denoted as $\widetilde{R}_B^{\ast}$ and $\widetilde{R}_E^{\ast}$, respectively. Our numerical results show that there is a unique pair of $\widetilde{R}_B^{\ast}$ and $\widetilde{R}_E^{\ast}$ that achieves a locally maximum $\widetilde{\Psi}_f(R_B, R_E, \rho_i, \rho_o)$ over all the anticipated operating conditions. One exemplary numerical result is presented in Fig.~\ref{fig:fixed_concave}. Substituting $\widetilde{R}_B^{\ast}$ and $\widetilde{R}_E^{\ast}$ into \eqref{throughput_fixed_threat}, we obtain the locally maximum value of $\widetilde{\Psi}_f(R_B, R_E, \rho_i, \rho_o)$, denoted by $\widetilde{\Psi}^{\ast}_f(\rho_i, \rho_o)$.

\subsection{Numerical Results}

In this subsection, we present numerical results to examine the impact of $\rho_i$ and $\rho_o$ on the determined wiretap code rates and the locally maximum average EST of the adaptive and fixed-rate transmission schemes.

\begin{figure}[!t]
    \begin{center}
    {\includegraphics[width=3.3in,height=2.8in]{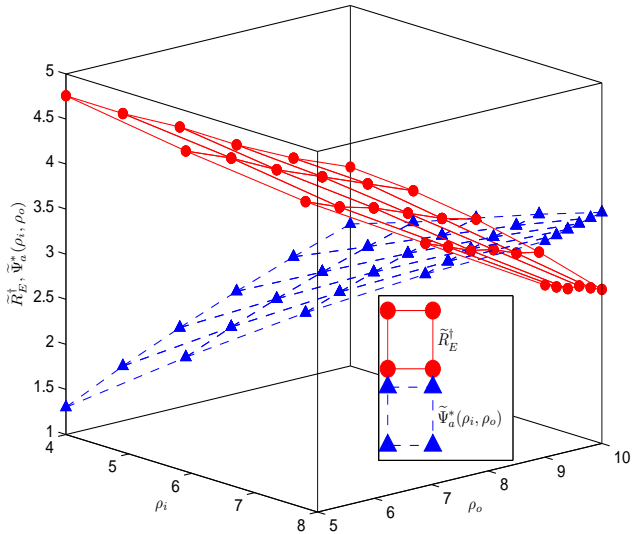}}
    \end{center}
    \caption{Redundancy rate, $R_{E, t}^{\dag}$, and the locally maximum average EST, $\widetilde{\Psi}^{\ast}_a(\rho_i, \rho_o)$, of the adaptive transmission scheme versus $\rho_i$ and $\rho_o$ for $N_E = 2$, ${\gamma}_B = 20$ dB, $\overline{\gamma}_0 = 30$ dB, $\rho_r = 1$, and $\eta = 3$.}\label{fig:threat_ada_3D}
\end{figure}

In Fig.~\ref{fig:threat_ada_3D}, we plot the redundancy rate, $\widetilde{R}_E^{\dag}$, and the locally maximum average EST, $\widetilde{\Psi}^{\ast}_a(\rho_i, \rho_o)$, of the adaptive transmission scheme versus $\rho_i$ and $\rho_o$. In this figure, we first observe that $\widetilde{R}_E^{\dag}$ decreases as $\rho_i$ increases, which reveals that the redundancy rate used to confuse Eve can be reduced by increasing the inner boundary. We also observe that $\widetilde{R}_E^{\dag}$ decreases as $\rho_o$ increases. This can be explained by the fact that a larger outer boundary $\rho_o$ means Eve is statistically further from Alice. Moreover, we observe $\widetilde{\Psi}^{\ast}_a(\rho_i, \rho_o)$ increases as $\rho_i$ increases, which indicates that the further the inner boundary is from Alice, the better secrecy performance the adaptive transmission scheme achieves. As such, a higher average EST can be achieved through enlarging the inner boundary. We further observe that $\widetilde{\Psi}^{\ast}_a(\rho_i, \rho_o)$ increases as $\rho_o$ increases. This can be explained by the fact that $\widetilde{\Psi}^{\ast}_a(\rho_i, \rho_o)$ is averaged over $\overline{\gamma}_E$ in the annulus threat model, and a larger $\rho_o$ means that Eve is further from Alice on average since in the annulus threat model Eve's location is uniformly distributed in the annulus.

\begin{figure}[!t]
    \begin{center}
    {\includegraphics[width=3.3in,height=2.8in]{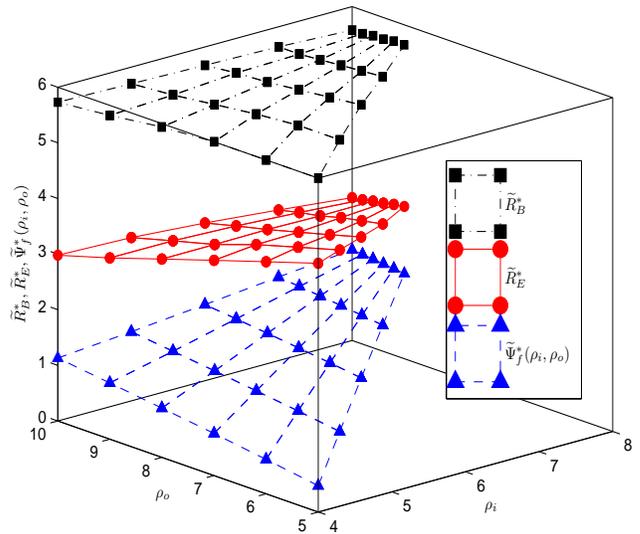}}
    \end{center}
    \caption{Wiretap code rates, $\widetilde{R}_B^{\ast}$ and $\widetilde{R}_E^{\ast}$, and the locally maximum average EST, $\widetilde{\Psi}^{\ast}_f(\rho_i, \rho_o)$, of the fixed-rate transmission scheme versus $\rho_i$ and $\rho_o$ for $N_E = 2$, $\overline{\gamma}_B = 20$ dB, $\overline{\gamma}_0 = 30$ dB, $\rho_r = 1$, and $\eta = 3$.}\label{fig:threat_fix_3D}
\end{figure}

In Fig.~\ref{fig:threat_fix_3D}, we plot the wiretap code rates, $\widetilde{R}_B^{\ast}$ and $\widetilde{R}_E^{\ast}$, and the locally maximum average EST, $\widetilde{\Psi}^{\ast}_f(\rho_i, \rho_o)$, of the fixed-rate transmission scheme versus $\rho_i$ and $\rho_o$. In this figure, we first observe that both $\widetilde{R}_B^{\ast}$ and $\widetilde{R}_E^{\ast}$ decrease as $\rho_i$ increases, and $\widetilde{R}_E^{\ast}$ is more sensitive to $\rho_i$ than $\widetilde{R}_B^{\ast}$, which results in $\widetilde{\Psi}^{\ast}_f(\rho_i, \rho_o)$ increasing as $\rho_i$ increases. The above observation indicates that the further the inner boundary is from Alice, the better secrecy performance the fixed-rate transmission scheme achieves. We also observe that both $\widetilde{R}_B^{\ast}$ and $\widetilde{R}_E^{\ast}$ increase as $\rho_o$ increases, and $\widetilde{R}_E^{\ast}$ is more sensitive to $\rho_o$ than $\widetilde{R}_B^{\ast}$, which results in  $\widetilde{\Psi}^{\ast}_f(\rho_i, \rho_o)$ increasing as $\rho_o$ increases. This is can be explained by the fact that $\widetilde{\Psi}^{\ast}_f(\rho_i, \rho_o)$ is averaged over $\overline{\gamma}_E$, and a larger $\rho_o$ means that Eve is statistically further from Alice.

We note that the issue of the unknown average SNR of
the eavesdropper's channel $\overline{\gamma}_E$ can be investigated via the consideration of a general average SNR model. In such an average SNR
model, a specific distribution (e.g., a uniform distribution or a
Gaussian distribution) can be assigned to $\overline{\gamma}_E$ and then thorough analysis of the average effective secrecy throughput of
the adaptive and fixed-rate transmission schemes can be conducted. We find that the specific distribution of
$\overline{\gamma}_E$ has to be carefully selected based on
practical application scenarios and corresponding system parameters.
This is due to the fact that the distribution 
of $\overline{\gamma}_E$ may depend on some specific system
parameters. For example, the distribution type of
$\overline{\gamma}_E$ is dependent on the path loss exponent $\eta$
within our annulus threat model.

\section{Conclusion}\label{conclusion}

In this paper, we proposed a new framework to determine the wiretap code rates for the SISOME wiretap channel. We considered  several scenarios including the case of the absolute-passive eavesdropping scenario in which even the average SNR of the eavesdropper's channel is unknown at the transmitter. The framework is based on a new performance metric, the EST, which
captures explicitly the reliability constraint and secrecy constraint of wiretap channels. The framework does not require a determination of the secrecy outage probability \emph{a priori} or subjectively, and therefore is very pragmatic.

\section*{Acknowledgement}

This work was funded by the Australian Research Council Grant DP120102607. We thank the three anonymous referees for their valuable comments and suggestions on improving the quality of the paper.



\section*{Appendix: Proof of Lemma 3}\label{proof_lemma3}

In the Cartesian coordinate system, we denote the location of Eve by $(u,v)$. In the annulus threat model, the joint pdf of $u$ and $v$ is
\begin{equation}\label{joint_pdf_uv}
f_{U,V}(u,v)=\left\{
\begin{array}{ll}
\frac{1}{\pi (\rho^2_o - \rho^2_i)}~\;, &  \mbox{$\rho_i^2 \leq u^2 + v^2 \leq \rho_o^2$,}\\
0~\;, & \mbox{otherwise}.\;
\end{array}
\right.
\end{equation}
In the polar coordinate system, we denote the location of Eve by $(\rho, \theta)$. As such, we obtain $u = \rho \cos \theta$ and $v = \rho \sin \theta$. The Jacobian matrix for this coordinate change is given by
\begin{align}\label{Jacobian_matrix}
J (\rho,\theta) = \frac{\partial (u,v)}{\partial (\rho,\theta)} = \left[\begin{array}{cc} \cos \theta & -\rho \sin \theta\\
\sin \theta  & \rho \cos \theta \end{array}\right].
\end{align}
Using \eqref{Jacobian_matrix}, the determinant of $J (\rho, \theta)$ is calculated as $|J (\rho, \theta)| = \rho$. Based on Jacobian techniques for the transformation of random variables \cite{leon1994probability}, the joint pdf of $\rho$ and $\theta$ is given by
\begin{equation}\label{joint_pdf_rt}
\begin{split}
f_{\Rho,\Theta}(\rho,\theta) &= f_{U,V}(u,v)|J (\rho,\theta)|\\
&= \left\{
\begin{array}{ll}
\frac{\rho}{\pi (\rho^2_o - \rho^2_i)}~\;, &  \mbox{$0 \!\leq\! \theta \!\leq\! 2 \pi$, $\rho_i \!\leq\! \rho \!\leq\! \rho_o$},\\
0~\;, & \mbox{otherwise}.\;
\end{array}
\right.
\end{split}
\end{equation}
Using \eqref{joint_pdf_rt}, the marginal pdf of $\rho$ is derived as
\begin{equation*}
f_{\Rho}(\rho) = \int_0^{2 \pi} f_{\Rho,\Theta}(\rho,\theta) d \theta
= \left\{
\begin{array}{ll}
\frac{2 \rho}{\rho^2_o - \rho^2_i}~\;, &\mbox{$\rho_i \!\leq\! \rho \!\leq\! \rho_o$},\\
0~\;, & \mbox{otherwise}.\;
\end{array}
\right.
\end{equation*}
In order to derive the pdf of $\rho^2$, we denote $\lambda = \rho^2$. As per the rules on the transformation of random variables, the pdf of $\rho^2$ is given by
\begin{equation}
f_{\Rho^2}(\rho^2) = \left|\frac{d \rho}{d \lambda}\right|f_{\Rho}(\rho) = \left\{
\begin{array}{ll}
\frac{1}{\rho^2_o - \rho^2_i}~\;, &\mbox{$\rho_i^2 \!\leq\! \rho^2 \!\leq\! \rho_o^2$},\\
0~\;, & \mbox{otherwise}.\;
\end{array}
\right.
\end{equation}
This completes the proof.


\begin{thebibliography}{1}

\bibitem{Shiu}
Y.-S. Shiu, S. Y. Chang, H.-C. Wu, S. C.-H. Huang, and H.-H. Chen,
``Physical layer security in wireless networks: A tutorial,''
\emph{IEEE Commun. Mag.}, vol. 18, no. 5, pp. 66--74, Apr. 2011.

\bibitem{mukherjee2011robust}
A. Mukherjee and A. L. Swindlehurst, ``Robust beamforming for
secrecy in MIMO wiretap channels with imperfect CSI,'' \emph{IEEE
Trans. Signal Process.}, vol. 59, no. 1, pp. 351--361, Jan. 2011.

\bibitem{zheng2011physical}
G. Zheng, P. Arapoglou, and B. Ottersten, ``Physical layer security
in multibeam satellite systems,'' \emph{IEEE Trans. Wireless
Commun.}, vol. 11, no. 2, pp. 852--863, Dec. 2011.

\bibitem{wang2013physical}
H. Wang, X. Zhou, and M. Reed, ``Physical layer security in cellular
networks: A stochastic geometry approach,'' \emph{IEEE Trans.
Wireless Commun.}, vol. 12, no. 6, pp. 2776--2787, May 2013.

\bibitem{geraci2014physical}
G. Geraci, H. S. Dhillon, J. G. Andrews, J. Yuan, and I. B. Collings, ``Physical Layer Security in Downlink Multi-Antenna Cellular Networks,'' \emph{IEEE Trans. Commun.}, vol. 62, no. 6, pp. 2006--2021, Jun. 2014.

\bibitem{shannon1949communication}
C. E. Shannon, ``Communication theory of secrecy systems,''
\emph{Bell Syst. Techn. J.}, vol. 28, no. 4, pp. 656--715, Oct.
1949.

\bibitem{wyner1975the}
A. Wyner, ``The wire-tap channel,'' \emph{Bell Syst. Techn. J.},
vol. 54, no. 8, pp. 1355--1387, Oct. 1975.

\bibitem{leung1978gaussian}
S. K. Leung-Yan-Cheong and M. E. Hellman, ``The Gaussian wire-tap
channel,'' \emph{IEEE Trans. Inf. Theory}, vol. 24, no. 4, pp.
451--456, Jul. 1978.

\bibitem{yan2014on} S. Yan, G. Geraci, N. Yang, R. Malaney, and J. Yuan, ``On the target secrecy rate for SISOME wiretap channels,'' in \emph{Proc. IEEE ICC}, Jun. 2014, pp. 987--992.

\bibitem{thangaraj2007applications}
A. Thangaraj, S. Dihidar, A. R. Calderbank, S. W. McLaughlin, and
J.-M. Merolla, ``Applications of LDPC codes to the wiretap
channel,'' \emph{IEEE Trans. Inf. Theory}, vol. 53, no. 8, pp.
2933--2945, Aug. 2007.


\bibitem{klinc2011ldpc} D. Klinc, J. Ha, S. W. McLaughlin, J. Barros, and B.-J. Kwak, ``LDPC codes for the Gaussian wiretap channel,'' \emph{IEEE Trans. Inf. Forensics Security}, vol. 6, no. 3, pp. 532--540, Sep. 2011.

\bibitem{harrison2013coding} W. K. Harrison, J. Almeida, M. R. Bloch, S. W. McLaughlin, and J. Barros, ``Coding for secrecy: an overview of error-control coding techniques for physical-layer security,'' \emph{IEEE Signal Process. Mag.}, vol. 30, no. 5, pp. 41--50, Sep. 2013.


\bibitem{gopala2008on} P. K. Gopala, L. Lai, and H. E. Gamal, ``On the secrecy capacity of fading channels,'' \emph{IEEE Trans. Inf. Theory}, vol. 54, no. 10, pp. 4687--4698, Oct. 2008.

\bibitem{shafiee2009towards} S. Shafiee, N. Liu, and S. Ulukus, ``Towards the secrecy capacity of the Gaussian MIMO wire-tap channel: The 2-2-1 Channel,'' \emph{IEEE Trans. Inf. Theory}, vol. 55, no. 9, pp. 4033--4039, Sep. 2009.

\bibitem{khisti2010secure2}
A. Khisti and G. W. Wornell, ``Secure transmission with multiple
antennas--Part II: The MIMOME wiretap channel,'' \emph{IEEE Trans.
Inf. Theory}, vol. 56, no. 11, pp. 5515--5532, Nov. 2010.


\bibitem{Oggier2011the}
F. Oggier and B. Hassibi,
``The secrecy capacity of the MIMO wiretap channel,''
\emph{IEEE Trans. Inf. Theory}, vol. 57, no. 8, pp. 4961--4972, Aug. 2011.




\bibitem{khisti2010secure} A. Khisti and G. W. Wornell, ``Secure transmission with multiple antennas I:
The MISOME wiretap channel,'' \emph{IEEE Trans. Inf. Theory}, vol. 56, no. 7, pp. 3088--3104, Jul. 2010.

\bibitem{parada2005secrecy} P. Parada and R. Blahut, ``Secrecy capacity of SIMO and slow fading
channels,'' in \emph{Proc. IEEE ISIT}, Sep. 2005, pp. 2152--2155.

\bibitem{bloch2008wireless} M. Bloch, J. Barros, M. Rodrigues, and S. McLaughlin, ``Wireless information-theoretic security,'' \emph{IEEE Trans. Inf. Theory}, vol. 54, no. 6, pp. 2515--2534, Jun. 2008.


\bibitem{alves2012performance}
H. Alves, R. D. Souza, M. Debbah, and M. Bennis, ``Performance of
transmit antenna selection physical layer security schemes,''
\emph{IEEE Signal Process. Lett.}, vol. 19, no. 6, pp. 372--375,
Jun. 2012.


\bibitem{yang2012transmit}
N. Yang, P. L. Yeoh, M. Elkashlan, R. Schober, and I. B. Collings,
``Transmit antenna selection for security enhancement in MIMO
wiretap channels,'' \emph{IEEE Trans. Commun.}, vol. 61, no. 1, pp.
144--154, Jan. 2013.

\bibitem{yan2013transmit} S. Yan, N. Yang, R. Malaney, and J. Yuan, ``Transmit antenna selection with Alamouti coding and power allocation in MIMO wiretap channels,'' \emph{IEEE Trans. Wireless Commun.}, vol. 13, no. 3, pp. 1656--1667, Mar. 2014.


\bibitem{shah2000performance} A. Shah and A.M. Haimovich, ``Performance analysis of maximal ratio combining and comparison with optimum combining for mobile radio communications with cochannel interference,'' \emph{IEEE Trans. Veh. Technol.}, vol. 49, no. 4, pp. 1454--1463, Jul. 2000.

\bibitem{chen2005analysis} Z. Chen, J. Yuan, and B. Vucetic, ``Analysis of transmit antenna selection/maximal-ratio combining in Rayleigh fading channels,'' \emph{IEEE Trans. Veh. Technol.}, vol. 54, no. 4, pp. 1312--1321, Jul. 2005.

\bibitem{goldsmith2005wireless} A. Goldsmith, \emph{Wireless Communications}, Cambridge, U.K.: Cambridge Univ. Press, 2005.

\bibitem{he2011maximal} F. He, H. Man, and W. Wang, ``Maximal ratio diversity combining enhanced security,'' \emph{IEEE Comm. Lett.}, vol. 15, no. 5, pp. 509--511, May 2011.

\bibitem{zhou2011rethinking} X. Zhou, M. R. McKay, B. Maham, and A. Hj{\o}rungnes, ``Rethinking the
secrecy outage formulation: A secure transmission design
perspective,'' \emph{IEEE Commun. Lett.}, vol. 15, no. 3, pp.
302--304, Mar. 2011.

\bibitem{abdallah2011keys} Y. Abdallah, M. A. Latif, M. Youssef, A. Sultan, and H. El Gamal, ``Keys through ARQ: Theory and practice,'' \emph{IEEE Trans. Inf. Forensics Security}, vol. 6, no. 3, pp. 737--751, Sep. 2011.

\bibitem{zhang2013on} X. Zhang, X. Zhou, and M. R. McKay, ``On the design of artificial-noise-aided secure multi-antenna transmission in slow fading channels,'' \emph{IEEE Trans. Veh. Technol.}, vol. 62, no. 5, pp. 2170--2181, Jun. 2013.

\bibitem{young1984orbits} N. J. Young, ``Orbits of the unit sphere of $\mathcal{L}(\mathcal{H}, \mathcal{K})$ under symplectic transformations,'' \emph{J. Operator Theory 11 (1984)}, pp. 171--191.

%



\bibitem{gradshteuin2007table}
I. S. Gradshteyn and I. M. Ryzhik, \emph{Table of Integrals, Series
and Products}, 7th ed., Academic, San Diego, CA, 2007.


\bibitem{leon1994probability} A. Leon-Garcia, \emph{Probability and Random Processes for Electrical Engineering}, 2nd ed. Cambridge, MA: Addison-Wesley, 1994.



%



\end{thebibliography}
\end{document}